\makeatletter
\declare@file@substitution{revtex4-1.cls}{revtex4-2.cls}
\makeatother

\documentclass[preprint]{aastex631}

\usepackage{natbib}
\usepackage{url}
\usepackage{xcolor}  
\usepackage{lineno}

\begin{document}

\title{The Pre-He White Dwarfs in Eclipsing Binaries. V. TIC 399725538 }
\correspondingauthor{Jae Woo Lee}
\email{jwlee@kasi.re.kr}
\author[0000-0002-5739-9804]{Jae Woo Lee}
\affil{Korea Astronomy and Space Science Institute, Daejeon 34055, Republic of Korea}

\author[0000-0002-8692-2588]{Kyeongsoo Hong}
\affil{Korea Astronomy and Space Science Institute, Daejeon 34055, Republic of Korea}

\author[0000-0002-8394-7237]{Min-Ji Jeong}
\affil{Korea Astronomy and Space Science Institute, Daejeon 34055, Republic of Korea}

\author[0000-0003-1916-9976]{Pakakaew Rittipruk}
\affil{National Astronomical Research Institute of Thailand, Chiang Mai 50200, Thailand}

\author[0000-0001-9339-4456]{Jang-Ho Park}
\affil{Korea Astronomy and Space Science Institute, Daejeon 34055, Republic of Korea}

\begin{abstract} 
We present echelle spectra of TIC 399725538 obtained in Korea and Thailand to investigate the physical properties and 
evolutionary scenarios of EL CVn-type binaries. The time-series spectra yielded the radial velocities (RVs) of 
the primary component and its atmospheric parameters, $T_{\rm eff,A}=7194\pm70$ K and $v_{\rm A}\sin i=68\pm9$ km s$^{-1}$. 
Joint modeling of our RVs and the archival TESS data provided component masses of $M_{\rm A}=1.930\pm0.054$ $M_\odot$ 
and $M_{\rm B}=0.211\pm0.005$ $M_\odot$, radii of $R_{\rm A}=1.922\pm0.020$ $R_\odot$ and $R_{\rm B}=0.207\pm0.005$ $R_\odot$, 
and luminosities of $L_{\rm A}=8.87\pm0.39$ $L_\odot$ and $L_{\rm B}=0.546\pm0.034$ $L_\odot$. The surface gravity and 
distance derived from this modeling are consistent with the model-independent $\log g_{\rm B}$ obtained from the direct 
observables and with the Gaia distance. The third light of $l_3=0.136\pm0.003$ comes mainly from two neighboring stars, 
TIC 399725539 and TIC 399725544. Comparison with stellar models indicates that TIC 399725538 A lies within 
the instability region of $\delta$ Sct$-$$\gamma$ Dor hybrids, whereas its extremely low-mass companion is markedly 
underluminous compared to theoretical white dwarf (WD) counterparts. Multifrequency analysis of the binary-subtracted 
residual lights revealed three significant signals, two of which correspond to aliases at two and four times 
the dominant frequency $f_1=0.0848$ day$^{-1}$. The $f_1$ frequency is likely a $\gamma$ Dor-type pulsation arising 
from the brighter A component, though further confirmation is required. Our results demonstrate that TIC 399725538 is 
a short-period EL CVn system belonging to the thick-disk population, consisting of a main-sequence $\gamma$ Dor pulsator 
and a helium-core WD precursor formed through stable mass transfer. 
\end{abstract}

\section{INTRODUCTION} 

The presumed scenario for extremely low-mass white dwarfs (ELM WDs, $M<0.3$ $M_\odot$) is that they are formed through 
non-conservative mass transfer $-$ where a significant fraction of the transferred mass is lost from the system rather than 
being accreted by the companion $-$ in binary stars, rather than from isolated single stars \citep{Marsh+1995,Kilic+2007,Chen+2017}. 
Most of their counterparts are intermediate-mass main-sequence stars, WDs, millisecond pulsars. EL CVn-type eclipsing binaries 
(EBs) are post-mass transfer systems containing an A/F primary dwarf \citep{Maxted+2014}, some of which exhibit pulsating signals 
in each component and are accompanied by circumbinary objects \citep[e.g.][]{Maxted+2013,Lagos+2020,Lee+2024}. Their study helps 
us understand the mass transfer and WD formation processes in binary systems. Although new EBs with ELM WD candidates are 
steadily being discovered by ground- and space-based missions \citep{Maxted+2014,vanRoestel+2018,Peng+2024,Xiong+2025}, 
their physical properties remain largely unknown due to a lack of good spectroscopic data. This paper continues our study for 
short-period EL CVn-type binaries, initiated with WASP 0131+28 \citep{Lee+2020}. The comprehensive reviews and main goals of 
this subject are detailed in \citet{Lee+2020,Lee+2024}. Here, we focus on TIC 399725538 (HD 287228, 2MASS J04530466+1015099, 
Gaia DR3 3294443634920790528; $T_{\rm p}$ = $+$10.810; $V$ = $+$11.258, $(B-V)$ = $+$0.402). 
 
The TESS target was added to our list of spectroscopic observations of EL CVn-type EBs with potential pulsations in 2023. 
It was first studied in detail by \citet{Peng+2024}. They searched for EL CVn systems in Gaia EBs \citep{Mowlavi+2023}, using 
archive data from TESS \citep{Ricker+2015} and Gaia \citep{Gaia2023}. This survey led to the discovery of four new candidates with 
typical light curves of the EL CVn type, one of which is TIC 399725538, the program target of this study. \citet{Peng+2024} modeled 
the TESS data observed at 1800-s and 600-s cadences, respectively, in Sectors 5 and 32, and combined the light curve parameters with 
Gaia spectroscopic elements to obtain the absolute dimensions of TIC 399725538. These results suggested that the EB system is 
a detached binary with a short period $P_{\rm orb}=1.29327$ days and an extreme mass ratio $q=0.077$, with the secondary component 
having the characteristics of a pre-ELM WD. Meanwhile, \citet{Peng+2024} applied Fourier analysis to the residual light curves 
obtained through their modeling, but no meaningful pulsation signal was found. This non-detection may be partly attributable to 
the limitations of the TESS data used in their analysis (e.g., undersampling).

In this article, we report the fundamental properties and possible pulsation signatures of TIC 399725538 based on in-depth analyses 
of our time-series echelle spectra and the recent TESS photometric data obtained from a higher-cadence mode. The brighter, massive 
component obscured in the secondary eclipse is denoted by the subscript A, and its hotter counterpart by the subscript B.

\section{TESS PHOTOMETRY AND ORBITAL EPHEMERIS}

\begin{figure}
\includegraphics[scale=0.9]{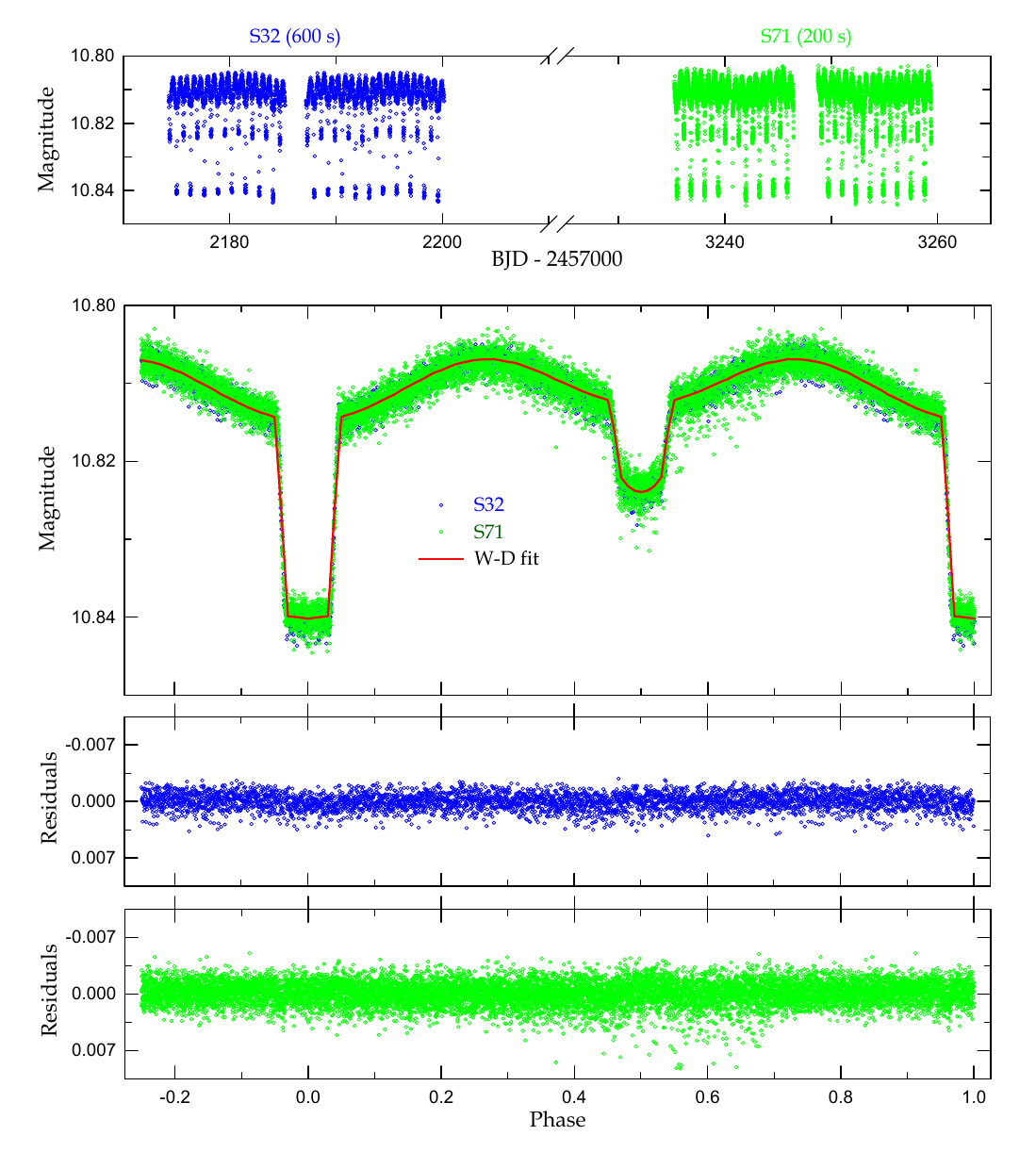}
\caption{TESS observations of TIC 399725538 distributed in BJD (top panel) and orbital phase (second panel). The blue and 
green circles are individual measurements for Sectors 32 and 71, respectively, and the red solid curve represents the synthetic 
model obtained through our W-D fit. The third and bottom panels show the residual lights corresponding to the binary model curve. }
\label{Fig1}
\end{figure}

The space-based light curves for TIC 399725538 are available in three observation sectors 5, 32, and 71 of the TESS mission 
\citep{Ricker+2015}. Sector 5 (S5) was observed at 1800-s cadence between 2018 November 15 and December 11, Sector 32 (S32) at 
600-s between 2020 November 20 and December 16, and Sector 71 (S71) at 200-s between 2023 October 16 and November 11. 
The time-series TESS data were collected from the MAST portal\footnote{\url{https://mast.stsci.edu/portal/Mashup/Clients/Mast/Portal.html}.  
The observations are available at \url{http://dx.doi.org/10.17909/c2qe-2177}.}. 
The S5 observations were excluded because their low sampling could introduce light-curve distortions such as amplitude suppression 
and frequency aliasing. We concentrated our analysis on the SAP data reduced from the SPOC pipeline \citep{Jenkins+2016}. 
The TESS observations had a significant interruption near the midpoint of each sector, allowing for data downlink. The raw flux 
data were detrended by introducing a linear least-squares fit to each segment split by this interruption \citep{Lee+2019,Lee+2024}. 
The corrected fluxes were transformed to magnitudes and are displayed in the top panel of Figure \ref{Fig1}.

The CROWDSAP values for S32 and S71 are 0.87049663 and 0.85590553, respectively. These mean that 86.3$\pm$0.7 \% of 
the observed fluxes in the target aperture originated from TIC 399725538, and the remainder could be contaminated by 
nearby sources. Two objects, TIC 399725539 ($T_{\rm p}$ = $+$14.01) and TIC 399725544 ($T_{\rm p}$ = $+$13.14), are located 
at distances of 16.33 and 23.08 arcsec from our target, respectively. Given the TESS CCD resolution of 21 arcsec pixel$^{-1}$, 
it is likely that the neighboring stars affected the TESS measurements. Their combined magnitude of 12.74 mag accounts for about 
14 \% of the total brightness of the three stars, which matches well with the 13.7 \% contamination from the CROWDSAP. 

To update the orbital ephemeris reported in \citet{Peng+2024}, we measured 67 minimum epochs from the TESS data \citep{Kwee+1956} 
as shown in Table \ref{Tab1} which resulted in the following ephemeris via least-squares fitting: 
\begin{equation}
 \mbox{Min I} = \mbox{BJD}~ 2,460,235.49338(\pm0.00048) + 1.29327809(\pm0.00000090)E.
\end{equation}
The $O-C$ timing residuals calculated using this ephemeris are presented in the fourth column of this table. The phase-folded 
light curves for S32 and S71 are plotted as blue and green circles, respectively, in the second panel of Figure \ref{Fig1}. 
The two eclipse types, Min I and Min II, are separated by a phase difference of 0.5, indicating that the binary star is probably 
in a circular orbit.

\section{ECHELLE SPECTROSCOPY AND SPECTRAL ANALYSIS}

We obtained a total of 33 echelle spectra of TIC 399725538 using two different facilities: the Bohyunsan Optical Astronomy 
Observatory (BOAO) in Korea and the Thai National Observatory (TNO) in Thailand. The BOAO observations were conducted over 
five nights, between 2023 November 10 and 2024 March 10, using the echelle spectrograph BOES \citep{Kim+2007} at the Doyak 
1.8-m telescope. Twenty-seven target spectra were acquired, each with an integral time of 1800 s, covering a wavelength range 
of 3600$-$10,200 \AA~ with a resolving power of $R \approx 30,000$. The remaining six spectra were taken on 2025 January 7 
using the TNT 2.4-m telescope and the MRES spectrograph \citep{Buisset+2018,Errmann+2020} at TNO, with the same exposure time 
of 1800 s. This spectrometer provides a spectral coverage of 3800$-$9000 \AA, with $R \approx 18,000$. 
The observed raw spectra were pre-processed using the \texttt{echelle} routines within \texttt{IRAF} in the same way as 
applied to WASP 0843–11 \citep{Hong+2021} and V389 Cas \citep{Rittipruk+2025}. The resulting spectra from both observatories 
exhibit a typical signal-to-noise ratio (SNR) of $\sim$15, each around 4500 \AA. 

\begin{figure}
\includegraphics[scale=1.0]{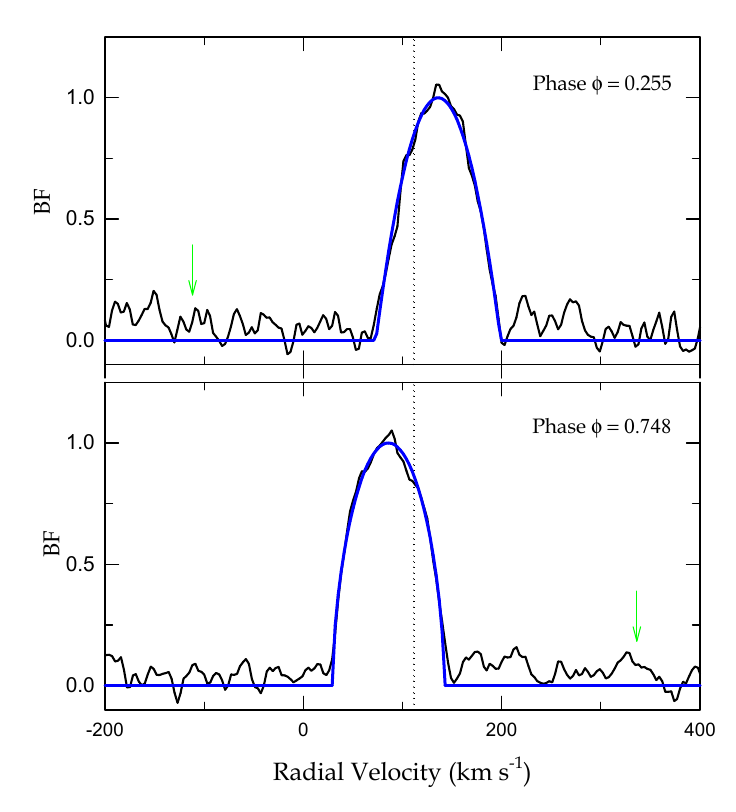}
\caption{Sample of the BF profiles. The black lines are the observed BFs for two orbital phases ($\phi$), with a single peak 
representing the primary component. A rotational broadening function applied to this peak is plotted as the solid blue line. 
The vertical dotted lines represent the radial velocity of the binary center of mass, and the green arrows point to the RV positions 
of the secondary companion predicted by our binary modeling. }
\label{Fig2}
\end{figure}

\begin{figure}
\includegraphics[scale=1.0]{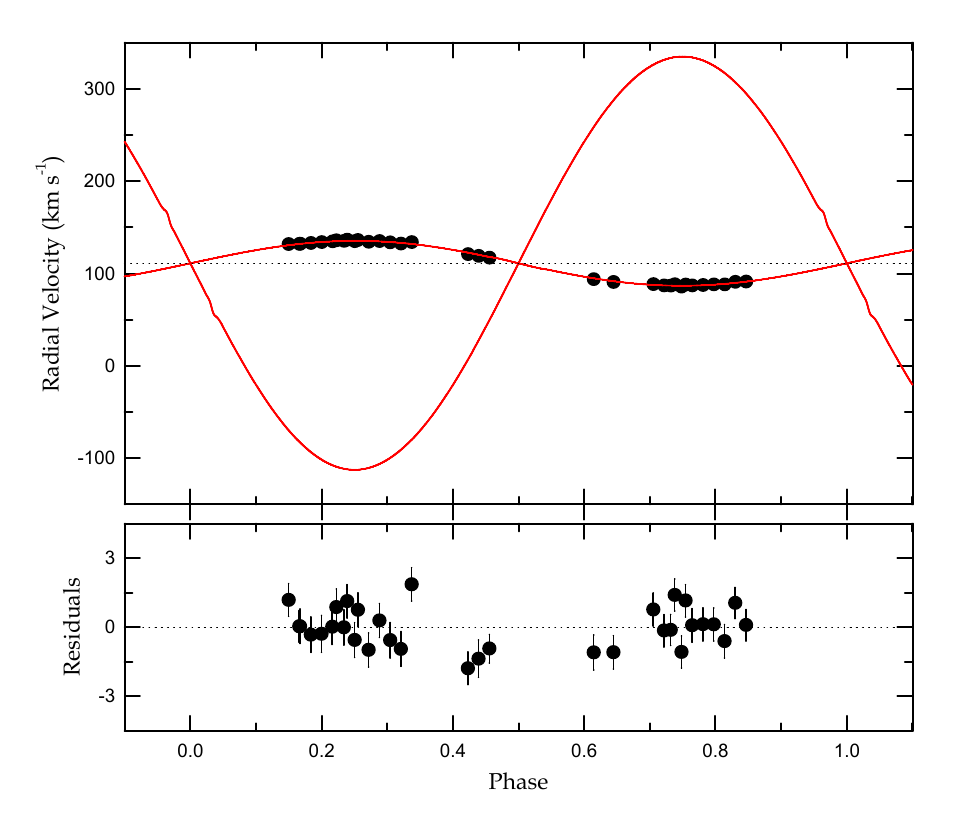}
\caption{Radial velocities of TIC 399725538 A. The solid curves represent the results from a consistent light and RV curve analysis, and 
the dotted line denotes the systemic velocity of $+$111.12 km s$^{-1}$. The lower panel displays the residuals between observations and models. }
\label{Fig3}
\end{figure}

We adopted the broadening function (BF) method \citep{Rucinski1992,Rucinski2002} implemented in the \texttt{RaveSpan} software 
\citep{Pilecki+2012} to yield the radial velocities (RVs) of TIC 399725538. The BF technique is particularly effective for 
resolving blended spectral lines in close binary systems. The synthetic template spectra were taken from the LTE library of 
\citet{Coelho+2005}. We focused the BF measurements on the spectral region of 4840$-$5085 \AA, where hydrogen and metallic lines, 
particularly the H$_{\rm \beta}$ and Mg \textsc{i} triplet, are prominent. This wavelength range was chosen to minimize 
contamination by strong telluric features. Figure \ref{Fig2} presents BF profiles taken at quadrature phases $\phi$ = 0.255 
(BJD 2,460,259.1027) and $\phi$ = 0.748 (BJD 2,460,317.9378). As seen in the samples, only a single dominant peak associated 
with the primary star is detected. The absence of companion signatures at the predicted RV positions, indicated by arrows in 
this figure, is likely due to the limited SNR of the observed spectra and the very low light contribution of the secondary component 
($\sim$2.3\%), which will be quantified through binary modeling in the next section. The RV measurements for TIC 399725538 A are 
summarized in Table \ref{Tab2} and plotted in Figure \ref{Fig3}. 

To reconstruct the representative spectrum of the primary component, we employed the \texttt{shift-and-add disentangling} method\footnote{\url{https://github.com/TomerShenar/Disentangling_Shift_And_Add/}} \citep{Shenar+2020, Shenar+2022}, 
optimized for the 3900$-$5085 \AA\ region of the observed target spectra. This spectral window includes the Balmer and 
metallic lines such as \ion{Ca}{2} K and H ($\lambda$3933.68 and $\lambda$3968.49), H$_{\rm \delta}$, H$_{\rm \gamma}$ 
and H$_{\rm \beta}$. The reconstructed spectrum was continuum-normalized using the neural-network-assisted \texttt{SUPPNET} 
algorithm\footnote{\url{https://github.com/RozanskiT/suppnet/}} \citep{Rozanski+2022}. The atmospheric modeling was then 
performed with the \texttt{GSSP} framework\footnote{\url{https://fys.kuleuven.be/ster/meetings/binary-2015/gssp-software-package}} 
\citep{Tkachenko2015}. In this approach, synthetic spectra were generated using \texttt{SynthV} \citep{Tsymbal1996}, 
based on stellar atmospheres from the \texttt{LLmodels} code \citep{Shulyak+2004}, and fitted to the reconstructed spectrum 
using a grid-search optimization method. 

\begin{figure}
\includegraphics[scale=0.9]{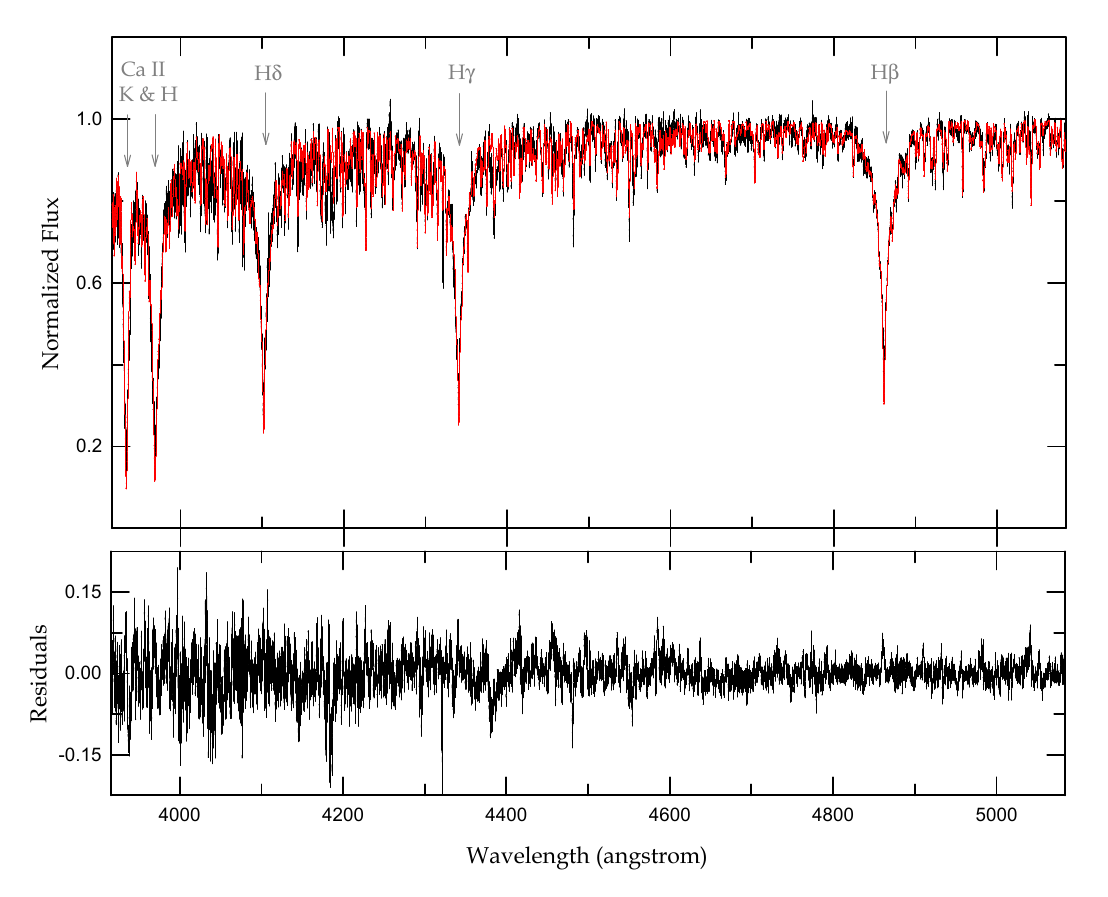}
\caption{Reconstructed spectrum of TIC 399725538 A. The black and red solid lines represent the disentangled and 
best-ﬁtting synthetic spectra, respectively. The lower panel displays the residuals between the two. }
\label{Fig4}
\end{figure}

The \texttt{GSSP} enables the simultaneous optimization of multiple atmospheric parameters: effective temperature 
($T_{\rm eff,A}$), surface gravity ($\log g_{\rm A}$), metallicity ([M/H]), projected rotational rate ($v_{\rm A}\sin i$), and 
turbulence velocities ($v_{\rm mic,A}$, $v_{\rm mac,A}$). The initial temperature and surface gravity of TIC 399725538 A was 
set at $T_{\rm eff,A}$ = $7398\pm83$ K and $\log g_{\rm A}$ = 4.0, as reported in \citet{Peng+2024}. Owing to the low SNR of 
our spectra, the solar metallicity was assumed. The microturbulence and macroturbulence velocities were estimated with 
the \texttt{iSpec}\footnote{\url{https://www.blancocuaresma.com/s/iSpec/}} \citep{Blanco-Cuaresma+2014} to be $v_{\rm mic,A}$ = 
1.91 km s$^{-1}$ and $v_{\rm mac,A}$ = 17.3 km s$^{-1}$, respectively. As the optimal result from the \texttt{GSSP} run, 
we found the atmospheric parameters of $T_{\rm eff,A}$ = 7194$\pm$70 K and $v_{\rm A}\sin i$ = 68$\pm$9 km s$^{-1}$ by fitting 
a polynomial to the minimum $\chi^2$ values \citep[cf.][]{Lehmann+2011}. Figure \ref{Fig4} shows the final synthetic model 
superimposed on the reconstructed spectrum.

\section{BINARY MODELING AND ABSOLUTE PARAMETERS}

The TESS light curve of TIC 399725538 shows the typical morphology of EL CVn-type EBs, with a box-shaped primary eclipse, 
a curved-bottom secondary eclipse, and an elliptical variation outside eclipse. To derive a consistent light and velocity solution 
for the program target, we simultaneously analyzed our ground-based RVs along with the space-based TESS data of S32 and S71 using 
the \texttt{Wilson-Devinney} (W-D) program, which is widely used to model EB observables \citep{Wilson+1971,Kallrath2022}. 
During this process, we did not use individual observations between BJD 2458252.7 and 2458253.5 exhibiting anomalous light variations, 
which we believe were due to instrumental issues. 
The mass ratio, $q$ = $M_{\rm B}/M_{\rm A}$ = $K_{\rm A}/K_{\rm B}$, is a key parameter that governs binary star modeling and 
determines fundamental properties such as mass, which is an essential factor in both star formation and stellar evolution. However, 
since there were no RV measurements for TIC 399725538 B, we could not directly determine its velocity semi-amplitude $K_{\rm B}$ 
and the corresponding $q$ value ​​from the present data. To estimate these parameters, we performed an extensive $q$-search procedure 
\citep[e.g.][]{Lee+2008,Lee+2022a}. 

To minimize degeneracy in this synthesis, we fixed several modeling parameters to reliable values ​​obtained from 
our spectroscopic analysis and different independent sources. The surface temperature of TIC 399725538 A was initially set at 
a central value of $T_{\rm A,eff}$ = 7194 K $-$ derived from our analysis of the disentangled spectrum $-$ to derive the best-fit solution. 
To account for the uncertainties propagated from this parameter, we performed additional analyses by setting $T_{\rm A,eff}$ to 
its lower (7124 K) and upper (7264 K) bounds (i.e., $\pm70$ K). A similar procedure was applied to the rotation-to-orbit velocity ratio 
of the more massive primary ($F_{\rm A}$), which was based on the observed $v_{\rm A}\sin i$ and the synchronous rotations 
$v_{\rm sync} = 2 \pi R_{\rm A}$/$P_{\rm orb}$. Specifically, $F_{\rm A}$ was fixed at 0.92 but varied by its error of $\pm0.12$ for 
uncertainty estimation, whereas TIC 399725538 B was assumed to be in synchronous rotation ($F_{\rm B}$ = 1.0). 

Because the temperatures of both components likely represent radiative envelopes, we adopted the standard values ​of $A_{\rm A,B}$ 
= 1.0 \citep{von1924} and $g_{\rm A,B}$ = 1.0 \citep{Rucinski1969} for the albedo and gravity-darkening parameters, respectively. 
Logarithmic limb-darkening coefficients ($x_{\rm A,B}$, $y_{\rm A,B}$) were taken from the updated tables of \citet{van1993} built 
in the W-D program. The circular orbit ($e$ = 0) of the binary target was adopted from the phase differences and the $O-C$ timing 
residuals of the two minimum types. The binary parameters set free during this modeling are the orbital ephemeris ($T_0$, $P_{\rm orb}$) 
and inclination ($i$), semi-major axis ($a$), center-of-mass velocity ($\gamma$), effective temperature of TIC 399725538 B ($T_{\rm B,eff}$), 
components' potentials ($\Omega _{\rm A,B}$), luminosity of TIC 399725538 A ($l_{\rm A}$), and the third light ($l_3$). 

We performed a $q$-search over the entire range of known mass ratios for the EL CVn binaries \citep{Lee+2022a}. 
Specifically, we computed a series of models for fixed $q$ values with a step size of 0.002, which was narrowed to 0.001 in 
the vicinity of the minimum. At each step, $q$ was held constant while other adjustable parameters were allowed to converge. 
The mass ratio is primarily constrained by the amplitude of the ellipsoidal  modulations in the outside-eclipse light curve, 
which arises from the tidal distortion of the components \citep{Morris1985,Bloemen+2012}. 
The weighted sum of squared residuals ($\sum W(O-C)^2$) exhibited a well-defined global minimum near $q = 0.11$. This value was 
subsequently adopted as the initial input for the final simultaneous solution where $q$ was adjusted as a free parameter. 
The resulting light and RV parameters are summarized in Table \ref{Tab3}. The binary star model indicates that TIC 399725538 
is a detached system with fill-out factors of $f_{\rm A}$ = 57.1 \% and $f_{\rm B}$ = 41.1 \%, where $f_{\rm A,B}$ = 
$\Omega_{\rm in}$/$\Omega_{\rm A,B}$ and $\Omega_{\rm in}$ represents the inner critical Roche potential. 
Most of the third light $l_3$ = 0.136 comes from two neighboring stars, TIC 399725539 and TIC 399725544, located around 
our program target, as discussed in the second section. Our synthetic model is presented as red solid curves in the second panel 
of Figure \ref{Fig1} and the top panel of Figure \ref{Fig3}, demonstrating an excellent fit to both the light and RV data. 

The simultaneous modeling parameters in Table \ref{Tab3} allow the calculation of the absolute dimensions of TIC 399725538, 
which are presented in Table \ref{Tab4} with those of \citet{Peng+2024}. In these computations, we used solar values of 
$M_{\rm bol}$$_\odot$ = +4.73 and $T_{\rm eff}$$_\odot$ = 5780 K, and obtained the bolometric corrections (BC) using 
the temperature-dependent calibration of each component \citep{Torres2010}. The secondary star parameters of 
$M_{\rm B}=0.211\pm0.005$ $M_\odot$, $R_{\rm B}=0.207\pm0.005$ $R_\odot$, $\log g_{\rm B} = 5.131\pm0.020$, and 
$T_{\rm B}$ = 10,935$\pm$110 K are typical values for pre-ELM WDs in EL CVn systems \citep{Maxted+2014,Lee+2020}. 
In single-lined EBs, the secondary's surface gravity can be calculated directly from observable quantities ($P_{\rm orb}$, 
$K_{\rm A}$, $e$, $r_{\rm B}$, $i$) using the following relation \citep{Southworth+2007}: 
\begin{equation}
g_B = {{2\pi} \over P_{\rm orb}} {{K_{\rm A}(1-e^2)^{1/2}} \over {r^2_{\rm B}\sin i}},  
\end{equation}
where $r_{\rm B}$ = $R_{\rm B}/a$ is the fractional radius of the low-mass companion. The surface gravity of $\log g_{\rm B}$ 
= 5.131$\pm$0.026 calculated from this equation is in perfect agreement with the value (5.131$\pm$0.020) derived from 
our binary modeling through $g_B = {{G M_{\rm B}} / {R^2_{\rm B}}}$. 

The target distance can be measured using the distance-modulus equation $V-A_{\rm V}-M_{\rm V,EB}=5\log{d}-5$. Here, $V$ is 
the apparent magnitude at maximum light, $A_{\rm V}$ $\simeq$ 3.1$E(B-V)$ is the interstellar extinction in $V$ band, and 
$M_{\rm V,EB}$ is the total absolute magnitude of the system, which is the sum of $M_{\rm V_{A}}$ and $M_{\rm V_{B}}$. 
Using the TESS v8.2 measurements of $V$ = 11.258$\pm$0.069 and $E$($B-V$) = 0.2381$\pm$0.0093 \citep{Paegert+2022}, 
we obtained a geometric distance of 477$\pm$19 pc for TIC 399725538. The EB-based distance matches well with the Gaia DR3 
distance of 484$\pm$6 pc, corresponding to a parallax of $\pi$ = 2.068$\pm$0.026 mas\citep{Gaia2023}.

\section{PULSATION FREQUENCY ANALYSIS}

\begin{figure}
\includegraphics[scale=1.0]{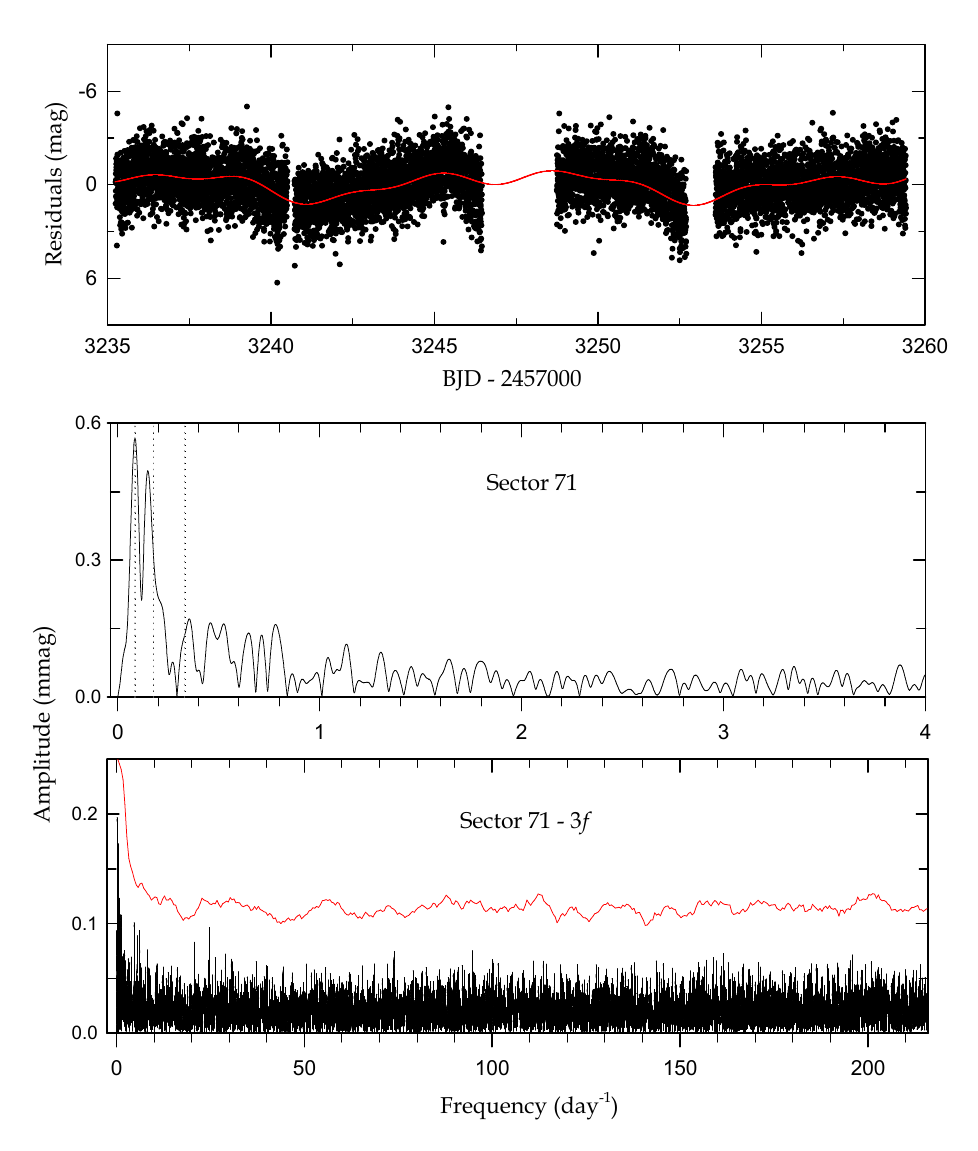}
\caption{The top panel shows the light curve residuals of TESS S71 distributed in BJD. The middle and bottom panels are amplitude spectra 
before and after prewhitening all three frequencies extracted from the full residuals using the \texttt{PERIOD04} code. The synthetic model 
curve for these frequencies is indicated by the red solid line in the top panel. The vertical dotted lines in the middle panel indicate the positions 
of the extracted frequencies and the red line in the bottom panel represents five times the noise spectrum. }
\label{Fig5}
\end{figure}

Considering the physical parameters in Table \ref{Tab4}, the primary and secondary components of TIC 399725538 can be classified as 
candidates for intermediate-mass and ELM WD pulsators, respectively \citep[e.g.,][]{Hong+2021,Lee+2022a,Lee+2024}. However, 
\citet{Peng+2024} reported no detectable pulsation signals in their TESS residual lights. We used the 200-s cadence data from S71 to 
find possible pulsation frequencies in the binary target. The top panel of Figure \ref{Fig5} shows the time-series light curve residuals 
from the W-D model fit. We applied the PERIOD04 software package \citep{Lenz+2005} to the binary-subtracted residuals, and searched 
for multiple frequencies up to the Nyquist frequency of 216 day$^{-1}$ through the pre-whitening process \citep{Lee+2014}. 

The results of the multi-frequency analysis are summarized in Table \ref{Tab5}, and the corresponding amplitude spectra are 
displayed in the middle and bottom panels of Figure \ref{Fig5}. We detected a total of three significant frequencies with SNR$>$5 \citep{Breger+1993,Baran+2021}, all of which lie within the $\gamma$ Dor instability region. The model light curve synthesized from 
these frequencies is indicated by a red solid line in the top panel of this figure. The total timespan of the TESS S71 data used is 
$\Delta T$ = 24.2 days. Using the Rayleigh criterion of $1/\Delta T = 0.041$ day$^{-1}$ as the frequency resolution for identifying 
aliasing signals, we found that the frequencies $f_2$ and $f_3$ correspond to approximately twice and four times $f_1$, respectively. 
After excluding these two harmonics, only the $f_1$ frequency is identified as a $\gamma$ Dor-type pulsation intrinsic to TIC 399725538 A. 

The TESS S71 residual lights exhibit gaps, including a data-download interruption between BJD 2,460,246.445 and BJD 2,460,248.741. 
Low-frequency artifacts may arise from uncorrected trends in TESS data or from imperfect removal of binary effects during the W-D 
modeling. Therefore, it is difficult to completely rule out the possibility that the $f_1$ frequency is an artifact. Furthermore, 
the 200-s sampling cadence is insufficient to detect short-period oscillations, such as those characteristic of ELM variables. 
More accurate short-cadence data will be required to confirm and refine our results through detailed pulsation analysis.

\section{DISCUSSION AND CONCLUSIONS}

For the TESS target TIC 399725538, classified as an EL CVn EB, we conducted time-series spectroscopy using the echelle spectrographs 
mounted on 2-m class telescopes in Korea and Thailand, respectively, securing a total of 33 spectra. We constructed BF profiles from 
the target spectra, showing only the peak corresponding to the primary component A. Despite the relatively low SNRs of the spectra, 
the absence of any detectable signal from the secondary companion indicates its extremely low light contribution ($l_{\rm B}$) and 
mass ratio ($q$). The RVs of TIC 399725538 A were derived by fitting the BF profiles with a rotational broadening function, and 
its atmospheric parameters were found to be $T_{\rm eff,A}$ = 7194$\pm$70 K and $v_{\rm A}\sin i$ = 68$\pm$9 km s$^{-1}$ by applying 
the \texttt{GSSP} package to a representative spectrum reconstructed using \texttt{shift-and-add disentangling} code. 
The ground-based spectroscopic measurements were combined with the time-series photometric data from TESS S32 and S71, yielding 
the following parameters for each component: $M_{\rm A}$ = 1.930$\pm$0.054 $M_\odot$, $R_{\rm A}$ = 1.922$\pm$0.020 $R_\odot$, 
$T_{\rm eff,A}$ = 7194$\pm$70 K, and $L_{\rm A}$ = 8.87$\pm$0.39 $L_\odot$ for TIC 399725538 A; and $M_{\rm B}$ = 0.211$\pm$0.005 $M_\odot$, 
$R_{\rm B}$ = 0.207$\pm$0.005 $R_\odot$, $T_{\rm eff,B}$ = 10,935$\pm$110 K, and $L_{\rm B}$ = 0.546$\pm$0.034 $L_\odot$ for TIC 399725538 B. 

\begin{figure}
\includegraphics[scale=1.0]{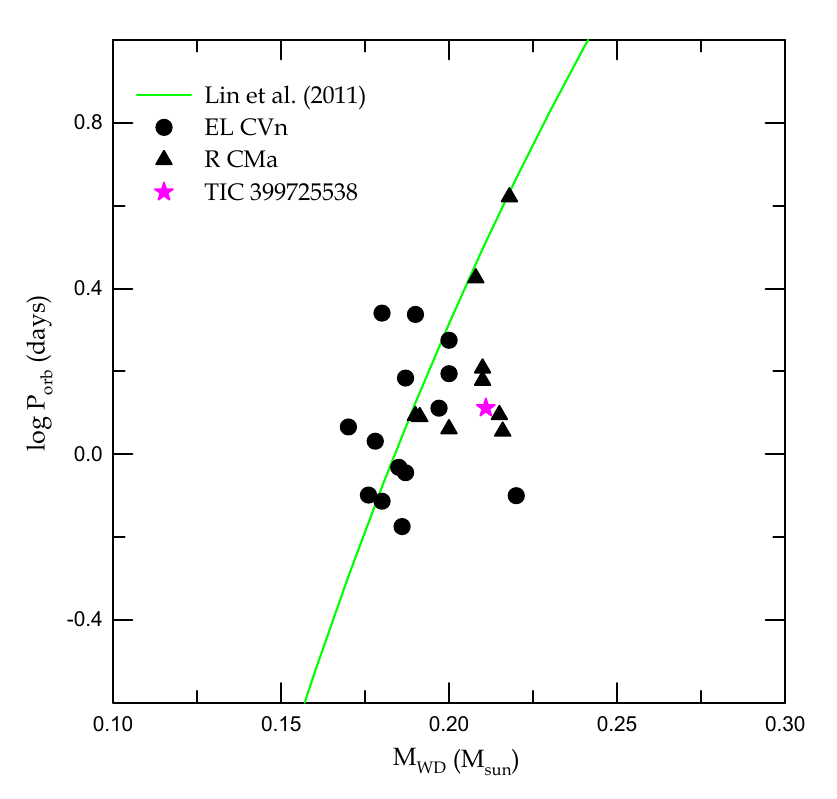}
\caption{$\log P_{\rm orb}-M_{\rm WD}$ diagram for TIC 399725538 B (star symbol) and the other pre-ELM WD companions in double-lined EL CVn-type 
(circles) and R CMa-type (triangles) binaries, respectively. The green solid line represents the relationship for the stable mass transfer 
evolution presented by \citet{Lin+2011}. } 
\label{Fig6}
\end{figure}

The absolute parameters indicate that TIC 399725538 is a typical EL CVn star, characterized by a short $P_{\rm orb}$ and low ($q$, $M_{\rm B}$) 
combination \citep{Maxted+2014,vanRoestel+2018,Lee+2020,Cakirli+2024}. If ELM WDs evolve through stable Roche-lobe overflow and 
their progenitors possess degenerate cores, a tight correlation would be expected between the He-core WD masses ($M_{\rm WD}$) and 
the orbital periods ($P_{\rm orb}$) \citep{Lin+2011,Chen+2017}. The position of TIC 399725538 B on the $\log P_{\rm orb}-M_{\rm WD}$ 
plot is shown in Figure \ref{Fig6}, along with the low-mass companions of short-period ($<$ 10 days) double-lined EL CVn and R CMa EBs \citep{Lee+Park2018,Lee+2018,Lee+2020,Lee+2022b,Lee+2024,Lee+2025,Wang+2019,Wang+2020,Kim+2021,Hong+2021,Cakirli+2024,Kovalev+2024}. 
In this figure, the pre-ELM WDs closely follow the $\log P_{\rm orb}-M_{\rm WD}$ relationship from \citet{Lin+2011}, supporting 
the prediction that the He-core WDs are formed in binary systems through non-conservative mass transfer \citep{Chen+2017}. 

\begin{figure}
\includegraphics[scale=1.0]{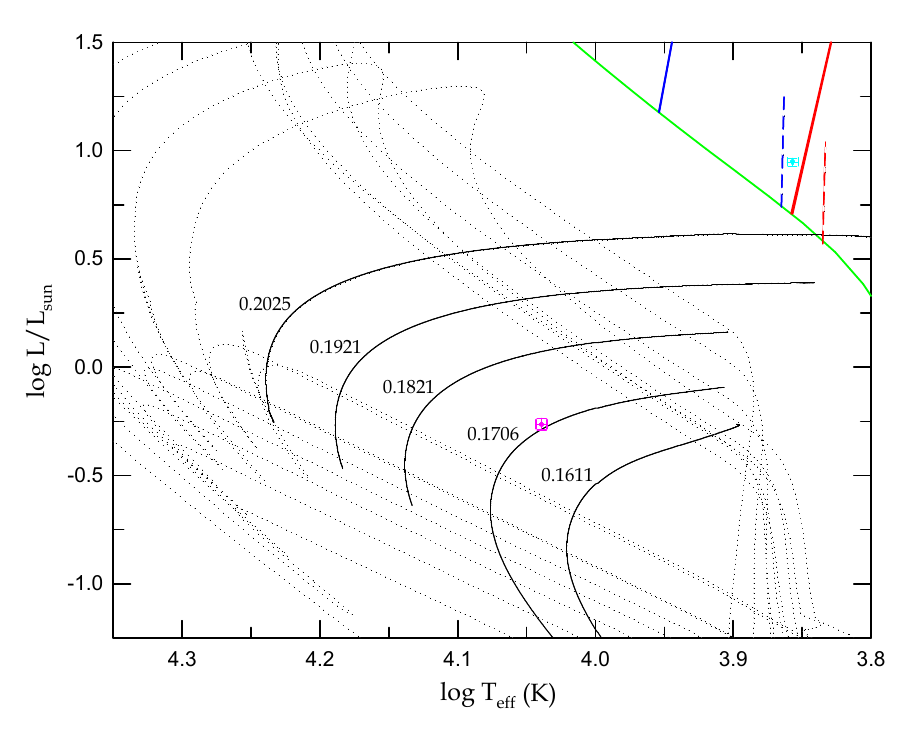}
\caption{Position of TIC 399725538 A (cyan cross) and B (pink cross) in the H-R diagram. The green solid line represents the zero-age main sequence, while 
the colored oblique solid and dashed lines indicate the instability strips of $\delta$ Sct and $\gamma$ Dor stars, respectively \citep[see][]{Lee+2024}. 
The black solid lines denote the evolutionary sequences of ELM WDs with masses from 0.1611 M$_\odot$ to 0.2025 M$_\odot$, up to the ﬁrst 
occurrence points of CNO ﬂashes \citep{Althaus+2013}. The dotted gray line shows the full evolutionary track of the 0.2025 M$_\odot$ model. } 
\label{Fig7}
\end{figure}

Comparing the binary parameters to stellar models provides a straightforward means of assessing whether observations are consistent 
with theoretical predictions. To examine the current evolutionary state of our program target, we employed 
the Hertzsprung-Russell (H-R) diagram presented in Figure \ref{Fig7}, where the cyan and pink crosses denote the primary (A) and 
secondary (B) components, respectively. TIC 399725538 A is located within the overlapping instability region of the main-sequence 
$\delta$ Sct and $\gamma$ Dor variables, suggesting that it is a candidate intermediate-mass pulsator. In the same figure, 
the solid black curves represent the evolutionary sequences for WD masses of 0.1611 $M_\odot$, 0.1706 $M_\odot$, 
0.1821 $M_\odot$, 0.1921 $M_\odot$, and 0.2025 $M_\odot$ from \citet{Althaus+2013}. The pre-ELM WD companion of TIC 399725538 
corresponds to an evolutionary track of 0.1706 $M_\odot$, which is about 19 \% less massive than our measured value. A similar trend 
is observed when employing the ELM WD models of \citet{Istrate+2016}, which account for both element diffusion and rotation mixing. 
These results indicate that TIC 399725538 B is significantly underluminous relative to its mass in the context of He-core WD evolution. 
The origin of the discrepancy between the target measurements and the ELM WD models remains unclear. However, it may stem from 
the lack of certain detailed physical processes in the WD formation models, or from unresolved binary evolution effects, such as 
non-conservative mass transfer. 

The population membership of WDs can be obtained through the classification scheme of \citet{Pauli+2006}, which is based on the $U-V$ 
and $J_{\rm z}-e$ diagrams. Following the procedure applied in \citet{Lee+2020,Lee+2024}, we calculated the space velocities and 
Galactic orbital parameters of TIC 399725538 as ($U, V, W$)\footnote{Positive toward the Galactic center, Galactic rotation, and 
North Galactic Pole.} = ($-86.6\pm0.2$, $+175.0\pm0.5$, $-33.0\pm0.1$) km s$^{-1}$ and ($J_{\rm z}$, $e$)\footnote{Angular momentum in 
the $z$ direction and eccentricity} = ($1484\pm18$ kpc km s$^{-1}$, $0.3282\pm0.0004$). These calculations used our systemic velocity 
$\gamma$ in conjunction with the Gaia measurements \citep{Gaia2023}. The location of TIC 399725538 in both diagrams suggests that 
the EL CVn EB is a member of the thick-disk population. This represents the third such system identified among 
well-studied EL CVn stars to date, following WASP 0247-25 \citep{Maxted+2014} and WASP 0346-21 \citep{Lee+2024}. 

Through the non-conservative mass transfer \citep{Chen+2017}, the initially more massive component of TIC 399725538 evolved 
into the present He-core WD precursor, while the mass gainer transformed into a potentially intermediate-mass, pulsating main-sequence star 
via accretion. TIC 399725538 B is currently in the pre-ELM WD stage, evolving toward higher $T_{\rm eff}$ at nearly constant $L$ 
before reaching a fully formed WD phase and subsequently entering the cooling sequence. Using the $M_{\rm WD}-t$ relation from \citet{Chen+2017}, 
we estimate its lifetime during this stage to be approximately 1.2 $\times$ 10$^{8}$ yr. 
Given that EL CVn-type binaries with reliable physical parameters remain rare, follow-up spectroscopic observations of poorly characterized 
targets are essential for elucidating the formation and evolution of these intriguing systems. In particular, high-resolution echelle 
spectroscopy using medium-to-large aperture telescopes, such as the VLT/UVES \citep[e.g.][]{Cakirli+2024,Lee+2024,Lee+2025}, will be crucial 
for detecting the low-mass companions and constraining their properties, especially considering the short orbital periods and high luminosity 
contrast that characterize EL CVn systems.

\begin{acknowledgements} 
This study utilizes echelle spectra obtained at the BOAO in Korea and the TNO in Thailand, together with light curves from 
the TESS mission. The TNO spectra were acquired with the MRES spectrograph under program ID TNTC12\_022. The authors thank 
the anonymous referee for providing insightful and constructive comments, which significantly improved the clarity and depth 
of the manuscript. We gratefully acknowledge support from the KASI grant 2026-1-904-01. 

\end{acknowledgements}

\clearpage

\clearpage 
\begin{deluxetable}{lcccc}
\tablewidth{0pt}
\tablecaption{TESS Eclipse Timings of TIC 399725538  \label{Tab1}}
\tablehead{
\colhead{BJD}    & \colhead{Error} & \colhead{Epoch} & \colhead{$O-C$} & \colhead{Min}  
}
\startdata
2,459,174.35735  & $\pm$0.00073    & -820.5          & $-0.00144$      & II            \\
2,459,175.00550  & $\pm$0.00026    & -820.0          & $+0.00007$      & I             \\
2,459,175.65159  & $\pm$0.00052    & -819.5          & $-0.00048$      & II            \\
2,459,176.29824  & $\pm$0.00038    & -819.0          & $-0.00047$      & I             \\
2,459,176.94625  & $\pm$0.00048    & -818.5          & $+0.00091$      & II            \\
2,459,177.59241  & $\pm$0.00021    & -818.0          & $+0.00043$      & I             \\
2,459,178.23721  & $\pm$0.00071    & -817.5          & $-0.00141$      & II            \\
2,459,178.88574  & $\pm$0.00026    & -817.0          & $+0.00048$      & I             \\
2,459,179.53301  & $\pm$0.00177    & -816.5          & $+0.00111$      & II            \\
2,459,180.17805  & $\pm$0.00027    & -816.0          & $-0.00049$      & I             \\
\enddata
\tablecomments{This table is available in its entirety in machine-readable form. A portion is shown here for guidance regarding its form and content.}
\end{deluxetable}

\begin{deluxetable}{lrcc}
\tablewidth{0pt}                    
\tabletypesize{\small}   
\tablecaption{Radial Velocities of TIC 399725538 A  \label{Tab2}}                                                                            
\tablehead{    
\colhead{BJD}          & \colhead{$V_{\rm A}$}   & \colhead{$\sigma_{\rm A}$}  & \colhead{Spectrograph}   \\                                            
\colhead{(2,460,000+)} & \colhead{(km s$^{-1}$)} & \colhead{(km s$^{-1}$)}     &    
} 
\startdata                                                                                            
259.0603               & $+135.87$               & 0.79                        & BOES                     \\
259.0815               & $+136.46$               & 0.72                        & BOES                     \\
259.1027               & $+136.16$               & 0.73                        & BOES                     \\
259.1238               & $+134.24$               & 0.74                        & BOES                     \\
259.1450               & $+135.09$               & 0.74                        & BOES                     \\
259.1662               & $+133.56$               & 0.77                        & BOES                     \\
259.1874               & $+132.26$               & 0.76                        & BOES                     \\
259.2086               & $+133.91$               & 0.72                        & BOES                     \\
259.3195               & $+120.88$               & 0.71                        & BOES                     \\
259.3407               & $+119.00$               & 0.81                        & BOES                     \\
259.3619               & $+117.07$               & 0.62                        & BOES                     \\
264.1393               & $+131.84$               & 0.71                        & BOES                     \\
264.1605               & $+132.06$               & 0.69                        & BOES                     \\
317.9167               & $+ 86.84$               & 0.68                        & BOES                     \\
317.9378               & $+ 85.75$               & 0.70                        & BOES                     \\
317.9590               & $+ 87.03$               & 0.73                        & BOES                     \\
317.9803               & $+ 87.45$               & 0.71                        & BOES                     \\
318.0015               & $+ 88.07$               & 0.71                        & BOES                     \\
318.0226               & $+ 88.20$               & 0.73                        & BOES                     \\
318.0438               & $+ 90.95$               & 0.68                        & BOES                     \\
318.0650               & $+ 91.31$               & 0.68                        & BOES                     \\
322.9381               & $+ 93.81$               & 0.76                        & BOES                     \\
322.9770               & $+ 90.69$               & 0.72                        & BOES                     \\
379.9596               & $+ 88.47$               & 0.71                        & BOES                     \\
379.9808               & $+ 87.02$               & 0.70                        & BOES                     \\
380.0020               & $+ 88.28$               & 0.70                        & BOES                     \\
380.0232               & $+ 88.00$               & 0.70                        & BOES                     \\
683.1844               & $+132.15$               & 0.75                        & MRES                     \\
683.2055               & $+132.92$               & 0.77                        & MRES                     \\
683.2267               & $+133.86$               & 0.81                        & MRES                     \\
683.2478               & $+134.83$               & 0.77                        & MRES                     \\
683.2707               & $+135.25$               & 0.76                        & MRES                     \\
683.2918               & $+134.85$               & 0.76                        & MRES                     \\
\enddata                                                                                                             
\end{deluxetable}

\begin{deluxetable}{lcc}
\tablewidth{0pt} 
\tablecaption{Light and Velocity Parameters of TIC 399725538  \label{Tab3}}
\tablehead{
\colhead{Parameter}               & \colhead{Primary (A)}  & \colhead{Secondary (B)}                                                  
}                                                                                                                                     
\startdata                                                                                                                            
$T_0$ (HJD)                       & \multicolumn{2}{c}{2,460,235.49365$\pm$0.00034}       \\
$P_{\rm orb}$ (day)               & \multicolumn{2}{c}{1.29327822$\pm$0.00000082}         \\
$a$ (R$_\odot$)                   & \multicolumn{2}{c}{6.436$\pm$0.058}                   \\
$\gamma$ (km s$^{-1}$)            & \multicolumn{2}{c}{$111.12\pm$0.24}                   \\
$K_{\rm A}$ (km s$^{-1}$)         & \multicolumn{2}{c}{24.41$\pm$0.37}                    \\
$K_{\rm B}$ (km s$^{-1}$)         & \multicolumn{2}{c}{223.7$\pm$2.2}                     \\
$q$                               & \multicolumn{2}{c}{0.1091$\pm$0.0020}                 \\
$i$ (deg)                         & \multicolumn{2}{c}{80.09$\pm$0.16}                    \\
$T_{\rm eff}$ (K)                 & 7194$\pm$70            & 10,935$\pm$110               \\
$\Omega$                          & 3.485$\pm$0.018        & 4.840$\pm$0.027              \\
$\Omega_{\rm in}$                 & \multicolumn{2}{c}{1.986}                             \\
$F$                               & 0.92$\pm$0.12          & 1.0                          \\   
$x$, $y$                          & 0.518, 0.284           & 0.390, 0.192                 \\
$l/(l_{\rm A}+l_{\rm B}+l_3)$     & 0.8415$\pm$0.0040      & 0.0225                       \\
$l_3$$\rm ^b$                     & \multicolumn{2}{c}{0.1359$\pm$0.0033}                 \\
$r$ (pole)                        & 0.2954$\pm$0.0016      & 0.0320$\pm$0.0007            \\
$r$ (point)                       & 0.3016$\pm$0.0018      & 0.0321$\pm$0.0007            \\
$r$ (side)                        & 0.2998$\pm$0.0017      & 0.0321$\pm$0.0007            \\
$r$ (back)                        & 0.3009$\pm$0.0018      & 0.0321$\pm$0.0007            \\
$r$ (volume)$\rm ^c$              & 0.2988$\pm$0.0017      & 0.0321$\pm$0.0007            \\
\enddata
\tablenotetext{\rm a}{Value at 0.25 orbital phase. }
\tablenotetext{\rm b}{Mean volume radius.}
\end{deluxetable}

\begin{deluxetable}{lccccc}
\tablewidth{0pt} 
\tablecaption{Absolute Parameters of TIC 399725538  \label{Tab4}}
\tablehead{
\colhead{Parameter}           & \multicolumn{2}{c}{\citet{Peng+2024}}        && \multicolumn{2}{c}{This Work}                     \\ [1.0mm] \cline{2-3} \cline{5-6} \\[-2.0ex]
                              & \colhead{Primary}   & \colhead{Secondary}    && \colhead{Primary}   & \colhead{Secondary}         
}                                                                                                                                                                                    
\startdata                                                                                                                                                                           
$M$ ($M_\odot$)               & 1.77$\pm$0.23       & 0.139$\pm$0.019        && 1.930$\pm$0.054     & 0.211$\pm$0.005             \\
$R$ ($R_\odot$)               & 2.14$\pm$0.11       & 0.237$\pm$0.013        && 1.922$\pm$0.020     & 0.207$\pm$0.005             \\
$\log$ $g$ (cgs)              & 4.024$\pm$0.029     & 4.827$\pm$0.056        && 4.156$\pm$0.010     & 5.131$\pm$0.020             \\
$\rho$ ($\rho_\odot$)         & \,                  & \,                     && 0.273$\pm$0.010     & 24.0$\pm$1.8                \\
$v_{\rm sync}$ (km s$^{-1}$)  & \,                  & \,                     && 75.18$\pm$0.80      & 8.08$\pm$0.19               \\
$v$$\sin$$i$ (km s$^{-1}$)    & \,                  & \,                     && 68$\pm$9            & \,                          \\
$T_{\rm eff}$ (K)             & 7398$\pm$83         & 10,740$\pm$120         && 7194$\pm$70         & 10,935$\pm$110              \\
$L$ ($L_\odot$)               & 12.3$\pm$1.8        & 0.67$\pm$0.10          && 8.87$\pm$0.39       & 0.546$\pm$0.034             \\
$M_{\rm bol}$ (mag)           & \,                  & \,                     && 2.362$\pm$0.048     & 5.387$\pm$0.067             \\
BC (mag)                      & \,                  & \,                     && 0.034$\pm$0.001     & $-$0.452$\pm$0.024          \\
$M_{\rm V}$ (mag)             & \,                  & \,                     && 2.328$\pm$0.048     & 5.839$\pm$0.072             \\
Distance (pc)                 & \,                  & \,                     && \multicolumn{2}{c}{477$\pm$19}                    \\
\enddata
\end{deluxetable}

\begin{deluxetable}{lccccc}
\tablewidth{0pt}
\tablecaption{Multi-frequency Analysis for TIC 399725538$\rm ^a$  \label{Tab5}}
\tablehead{
             & \colhead{Frequency}    & \colhead{Amplitude} & \colhead{Phase} & \colhead{SNR$\rm ^b$} & \colhead{Remark}      \\
             & \colhead{(day$^{-1}$)} & \colhead{(mmag)}    & \colhead{(rad)} &                       & 
}
\startdata
$f_{1}$      & 0.0848$\pm$0.0010      & 0.695$\pm$0.087     & 5.47$\pm$0.37   & 13.73                 &                       \\
$f_{2}$      & 0.1758$\pm$0.0017      & 0.427$\pm$0.086     & 2.49$\pm$0.59   &  8.50                 & 2$f_1$                \\
$f_{3}$      & 0.3310$\pm$0.0026      & 0.269$\pm$0.085     & 3.07$\pm$0.93   &  5.39                 & 4$f_1$                \\
\enddata                                                                                                                           
\tablenotetext{\rm a}{Parameter errors were calculated following \citet{Kallinger+2008}. }
\tablenotetext{\rm b}{Calculated in a range of 5 day$^{-1}$ around each frequency. }
\end{deluxetable}


\newpage
\begin{thebibliography}{}
\providecommand{\dodoi}[1]{doi:~\href{http://doi.org/#1}{\nolinkurl{#1}}}
\providecommand{\doarXiv}[1]{\href{https://arxiv.org/abs/#1}{\nolinkurl{https://arxiv.org/abs/#1}}}

\bibitem[L. G. Althaus et al.(2013)]{Althaus+2013} Althaus, L. G., Miller Bertolami, M. M., \& C\'orsico, A. H. 2013, A\&A, 557, A19, \dodoi{ 
10.1051/0004-6361/201321868} 
\bibitem[A. S. Baran \& C. Koen(2021)]{Baran+2021} Baran, A. S., \& Koen, C. 2021, AcA, 71, 113, \dodoi{10.32023/0001-5237/71.2.3} 
\bibitem[S. Blanco-Cuaresma et al.(2014)]{Blanco-Cuaresma+2014} Blanco-Cuaresma, S., Soubiran, C., Heiter, U., et al. 2014, A\&A, 569, A111, \dodoi{10.1051/0004-6361/201423945} 
\bibitem[S. Bloemen et al.(2012)]{Bloemen+2012} Bloemen, S., Marsh, T. R., Degroote, P., et al. 2012, MNRAS, 422, 2600, \dodoi{10.1111/j.1365-2966.2012.20818.x} 
\bibitem[M. Breger et al.(1993)]{Breger+1993} Breger, M., Stich, J., Garrido, R., et al. 1993, A\&A, 271, 482, \dodoi{10.1051/0004-6361:1993271482}
\bibitem[C. Buisset et al.(2018)]{Buisset+2018} Buisset, C., Poshyachinda, S., Soonthornthum, B., et al. 2018, Proc. SPIE,\dodoi{10.1117/12.2299493} 
\bibitem[\" O. \c Cakirli et al.(2024)]{Cakirli+2024} \c Cakirli, \" O., Hoyman, B., \& \" Ozdarcan, O. 2024, MNRAS, 533, 2058 \dodoi{10.1093/mnras/stae1948} 
\bibitem[X. Chen et al.(2017)]{Chen+2017} Chen, X., Maxted, P. F. L., Li, J., \& Han, Z. 2017, MNRAS, 467, 1874, \dodoi{10.1093/mnras/stx115} 
\bibitem[P. Coelho et al.(2005)]{Coelho+2005} Coelho, P., Barbuy, B., Melendez, J., Sciavon, R. P., \& Castilho, B. V. 2005, A\&A, 443, 735, \dodoi{10.1051/0004-6361:20053511} 
\bibitem[R. Errmann et al.(2020)]{Errmann+2020} Errmann, R., Cook, N., Anglada-Escudé, G., et al. 2020, PASP, 132, 064504,  
\dodoi{10.1088/1538-3873/ab8783} 
\bibitem[Gaia Collaboration et al.(2023)]{Gaia2023} Gaia Collaboration, Vallenari A., Brown A. G. A., et al. 2023, A\&A, 674, A1, \dodoi{10.1051/0004-6361/202243940} 
\bibitem[K. Hong et al.(2021)]{Hong+2021} Hong, K., Lee, J. W., Koo, J.-R., et al. 2021, AJ, 161, 137, \dodoi{10.3847/1538-3881/abdd39} 
\bibitem[A. G. Istrate et al.(2016)]{Istrate+2016} Istrate, A. G., Marchant, P., Tauris, T. M., et al. 2016, A\&A, 595, A35, \dodoi{10.1051/0004-6361/201628874} 
\bibitem[J. M. Jenkins et al.(2016)]{Jenkins+2016} Jenkins, J. M., Twicken, J. D., McCauliff, S., et al. 2016, Proc. SPIE, 9913, 99133E, \dodoi{10.1117/12.2233418} 
\bibitem[T. Kallinger et al.(2008)]{Kallinger+2008} Kallinger, T., Reegen, P., \& Weiss, W. W. 2008, A\&A, 481, 571, \dodoi{10.1051/0004-6361:20077559}
\bibitem[J. Kallrath(2022)]{Kallrath2022} Kallrath, J. 2022, Galaxies, 10, 17, \dodoi{10.3390/galaxies10010017} 
\bibitem[M. Kilic et al.(2007)]{Kilic+2007} Kilic, M., Stanek, K. Z., \& Pinsonneault, M. H. 2007, ApJ, 671, 761, \dodoi{10.1086/522228}
\bibitem[K.-M. Kim et al.(2007)]{Kim+2007} Kim, K.-M., Han, I., Valyavin, G. G., et al. 2007, PASP, 119, 1052, \dodoi{10.1086/521959} 
\bibitem[S.-L. Kim et al.(2021)]{Kim+2021} Kim, S.-L., Lee, J. W., Lee, C.-U., et al. 2021, AJ, 162, 212, \dodoi{10.3847/1538-3881/ac23de} 
\bibitem[M. Kovalev et al.(2024)]{Kovalev+2024} Kovalev, M., Li, Z., Xiong, J., et al. 2024, MNRAS, 535, 2651, \dodoi{10.1093/mnras/stae2494} 
\bibitem[K. K. Kwee \& H. Van Woerden(1956)]{Kwee+1956} Kwee, K. K., \& Van Woerden, H. 1956, Bull. Astron. Inst. Netherlands, 12, 327
\bibitem[F. Lagos et al.(2020)]{Lagos+2020} Lagos, F., Schreiber, M. R., Parsons, S. G., et al. 2020, MNRAS, 499, L121, \dodoi{10.1093/mnrasl/slaa164}
\bibitem[J. W. Lee \& J.-H. Park(2018)]{Lee+Park2018} Lee, J. W., \& Park, J.-H. 2018, MNRAS, 480, 4693, \dodoi{10.1093/mnras/sty2153} 
\bibitem[J. W. Lee et al.(2024)]{Lee+2024} Lee, J. W., Hong, K., Jeong, M.-J., \& Wolf, M. 2024, ApJ, 973, 113, \dodoi{10.3847/1538-4357/ad67c7} 
\bibitem[J. W. Lee et al.(2022a)]{Lee+2022a} Lee, J. W., Hong, K., Kim, H.-Y., \& Park, J.-H. 2022a, MNRAS, 515, 4702, \dodoi{10.1093/mnras/stac2151} 
\bibitem[J. W. Lee et al.(2018)]{Lee+2018} Lee, J. W., Hong, K., Koo, J.-R., \& Park J.-H., 2018, AJ, 155, 5, \dodoi{10.3847/1538-3881/aa947e} 
\bibitem[J. W. Lee et al.(2022b)]{Lee+2022b} Lee, J. W., Hong, K., \& Park, J.-H. 2022b, MNRAS, 511, 654, \dodoi{10.1093/mnras/stac075} 
\bibitem[J. W. Lee et al.(2025)]{Lee+2025} Lee, J. W., Jeong, M.-J., \& Hong, K. 2025, MNRAS, 538, 3314, \dodoi{10.1093/mnras/staf473} 
\bibitem[J. W. Lee et al.(2014)]{Lee+2014} Lee, J. W., Kim, S.-L., Hong, K., Lee, C.-U., \& Koo, J.-R. 2014, AJ, 148, 37, \dodoi{10.1088/0004-6256/148/2/37} 
\bibitem[J. W. Lee et al.(2020)]{Lee+2020} Lee, J. W., Koo, J.-R., Hong, K., \& Park, J.-H. 2020, AJ, 160, 49, \dodoi{10.3847/1538-3881/ab9621} 
\bibitem[J. W. Lee et al.(2019)]{Lee+2019} Lee, J. W., Kristiansen, M., \& Hong, K. 2019, AJ, 157, 223, \dodoi{10.3847/1538-3881/ab1a3b} 
\bibitem[J. W. Lee et al.(2008)]{Lee+2008} Lee, J. W., Youn, J.-H., Kim, C.-H., Lee, C.-U., \& Kim, H.-I. 2008, AJ, 135, 1523, \dodoi{10.1088/0004-6256/135/4/1523} 
\bibitem[H. Lehmann et al.(2011)]{Lehmann+2011} Lehmann, H., Tkachenko, A., Semaan, T., et al. 2011, A\&A, 526, A124, \dodoi{10.1051/0004-6361/201015769} 
\bibitem[P. Lenz \& M. Breger(2005)]{Lenz+2005} Lenz, P., \& Breger, M. 2005, Comm. Asteroseismology, 146, 53, \dodoi{10.1553/cia146s53} 
\bibitem[J. Lin et al.(2011)]{Lin+2011} Lin, J., Rappaport, S., Podsiadlowski, P., et al. 2011, ApJ, 732, 70, \dodoi{10.1088/0004-637X/732/2/70} 
\bibitem[T. R. Marsh et al.(1995)]{Marsh+1995} Marsh, T. R., Dhillon, V. S., \& Duck, S. R. 1995, MNRAS, 275, 828, \dodoi{10.1093/mnras/275.3.828} 
\bibitem[P. F. L. Maxted et al.(2014)]{Maxted+2014} Maxted, P. F. L., Bloemen, S., Heber, U., et al. 2014, MNRAS, 437, 1681, \dodoi{10.1093/mnras/stt2007} 
\bibitem[P. F. L. Maxted et al.(2013)]{Maxted+2013} Maxted, P. F. L., Serenelli, A. M., Miglio, A., et al. 2013, Natur, 498, 463, \dodoi{10.1038/nature12192}
\bibitem[S. L. Morris(1985)]{Morris1985} Morris, S. L. 1985, ApJ, 295, 143, \dodoi{10.1086/163359} 
\bibitem[N. Mowlavi et al.(2023)]{Mowlavi+2023} Mowlavi, N., Holl, B., Lecoeur-Ta\" ïbi, I., et al. 2023, A\&A, 674, A16, \dodoi{10.1051/0004-6361/202245330} 
\bibitem[M. Paegert et al.(2022)]{Paegert+2022} Paegert, M., Stassun, K. G., Collins, K. A., et al. 2022, VizieR Online Data Catalog, IV/39 
\bibitem[E.-M. Pauli et al.(2006)]{Pauli+2006} Pauli, E.-M., Napiwotzki, R., Heber, U., Altmann, M., \& Odenkirchen, M. 2006, A\&A, 447, 173, \dodoi{10.1051/0004-6361:20052730} 
\bibitem[Y. Peng et al.(2024)]{Peng+2024} Peng, Y., Wang, K., Ren, A., 2024, NewA, 107, 102153, \dodoi{10.1016/j.newast.2023.102153} 
\bibitem[B. Pilecki et al.(2012)]{Pilecki+2012} Pilecki, B., Konorski, P., \& Gorski, M. 2012, in IAU Symp. 282, From Interacting Binaries to Exoplanets: Essential Modeling Tools, ed. M. T. Richards, \& I. Hubeny (Cambridge: Cambridge Univ. Press), 301, \dodoi{10.1017/S174392131102761X} 
\bibitem[G. R. Ricker et al.(2015)]{Ricker+2015} Ricker, G. R., Winn, J. N., Vanderspek, R., et al. 2015, JATIS, 1, 014003, \dodoi{10.1117/1.JATIS.1.1.014003} 
\bibitem[P. Rittipruk et al.(2025)]{Rittipruk+2025} Rittipruk, P., Hong, K., Lee, J. W., et al. 2025, AJ, 169, 66, \dodoi{10.3847/1538-3881/ad99ce} 
\bibitem[T. R{\'o}{\.z}a{\'n}ski et al.(2022)]{Rozanski+2022} R{\'o}{\.z}a{\'n}ski, T., Niemczura, E., Lemiesz, J., et al. 2022, A\&A, 659, A199, \dodoi{10.1051/0004-6361/202141480} 
\bibitem[S. M. Rucinski(1969)]{Rucinski1969} Rucinski, S. M. 1969, Acta. Astron., 19, 125
\bibitem[S. M. Rucinski(1992)]{Rucinski1992} Rucinski, S. M. 1992, AJ, 104, 1968, \dodoi{10.1051/0004-6361/201015769} 
\bibitem[S. M. Rucinski(2002)]{Rucinski2002} Rucinski, S. M. 2002, AJ, 124, 1746, \dodoi{10.1086/342342} 
\bibitem[T. Shenar et al.(2020)]{Shenar+2020} Shenar, T., Bodensteiner, J., Abdul-Masih, M., et al. 2020, A\&A, 639, L6, \dodoi{10.1051/0004-6361/202038275} 
\bibitem[T. Shenar et al.(2022)]{Shenar+2022} Shenar, T., Sana, H., Mahy, L., et al. 2022, A\&A, 665, A148, \dodoi{10.1051/0004-6361/202244245} 
\bibitem[D. Shulyak et al.(2004)]{Shulyak+2004} Shulyak, D., Tsymbal, V., Ryabchikova, T., et al. 2004, A\&A, 428, 993, \dodoi{10.1051/0004-6361:20034169} 
\bibitem[J. Southworth et al.(2007)]{Southworth+2007} Southworth, J., Wheatley, P. J., Sams, G. 2007, MNRAS, 379, L11, \dodoi{10.1111/j.1745-3933.2007.00324.x} 
\bibitem[A. Tkachenko(2015)]{Tkachenko2015} Tkachenko, A. 2015, A\&A, 581, A129, \dodoi{10.1051/0004-6361/201526513} 
\bibitem[G. Torres(2010)]{Torres2010} Torres, G. 2010, AJ, 140, 1158, \dodoi{10.1088/0004-6256/140/5/1158} 
\bibitem[V. Tsymbal(1996)]{Tsymbal1996} Tsymbal, V. 1996, in ASP Conf. Ser. 108, M.A.S.S., Model Atmospheres and Spectrum Synthesis, ed. S. J. Adelman, F. Kupka, \& W. W. Weiss (San Francisco, CA: ASP), 198
\bibitem[W. Van Hamme(1993)]{van1993} Van Hamme, W. 1993, AJ, 106, 209, \dodoi{10.1086/116788} 
\bibitem[J. van Roestel et al. (2018)]{vanRoestel+2018} van Roestel, J., Kupfer, T., Ruiz-Carmona, R., et al. 2018, MNRAS, 475, 2560, \dodoi{ 
10.1093/mnras/stx3291} 
\bibitem[H. Von Zeipel(1924)]{von1924} Von Zeipel, H., 1924, MNRAS, 84, 665, \dodoi{10.1093/mnras/84.9.665} 
\bibitem[K. Wang et al.(2019)]{Wang+2019} Wang, K., Zhang, X., Luo, Y., Luo, C. 2019, MNRAS, 486, 2462, \dodoi{10.1093/mnras/stz1033} 
\bibitem[L. Wang et al.(2020)]{Wang+2020} Wang, L., Gies, D. R., Lester, K. V., et al. 2020, AJ, 159, 4, \dodoi{10.3847/1538-3881/ab52fa} 
\bibitem[R. E. Wilson \& E. J. Devinney(1971)]{Wilson+1971} Wilson, R. E., \& Devinney, E. J. 1971, ApJ, 166, 605, \dodoi{10.1086/150986} 
\bibitem[J. Xiong et al.(2025)]{Xiong+2025} Xiong, J., Li, Z., Li, J., et al. 2025, ApJ, 979, 108, \dodoi{10.3847/1538-4357/ad9b9c} 
\end{thebibliography}
\end{document}